# Frequency-doubled femtosecond Er-doped fiber laser for two-photon excited fluorescence imaging

Dorota Stachowiak,[1,†] Jakub Bogusławski,[2,†] Aleksander Głuszek,[1] Zbigniew Łaszczych,[1] Maciej Wojtkowski,[2] and Grzegorz Soboń[1,*]

[1]*Laser & Fiber Electronics Group, Faculty of Electronics, Wrocław University of Science and Technology, Wybrzeże Wyspianskiego 27, 50-370 Wrocław, Poland*
[2]*Institute of Physical Chemistry, Polish Academy of Sciences, Kasprzaka 44/52, 01-224 Warsaw, Poland*
[†]*These Authors contributed equally to this work*
*\*grzegorz.sobon@pwr.edu.pl*

**Abstract:** A femtosecond frequency-doubled Erbium-doped fiber laser with adjustable pulse repetition rate is developed and applied in two-photon excited fluorescence microscopy. The all-fiber laser system provides the fundamental pulse at 1560 nm wavelength with 22 fs duration for the second harmonic generation, resulting in 1.35 nJ, 60 fs pulses at 780 nm. The repetition rate is adjusted by a pulse picker unit built-in within the amplifier chain, directly providing transform-limited pulses for any chosen repetition rate between 1 and 12 MHz. We employed the laser source to drive a scanning two-photon excited fluorescence microscope for *ex vivo* rat skin and other samples' imaging at various pulse repetition rates. Due to compactness, ease of operation, and suitable pulse characteristics the laser source can be considered as an attractive alternative for Ti:sapphire laser in biomedical imaging.

## 1. Introduction

Femtosecond lasers are the enabling technology in the field of biophotonics, including non-invasive diagnostics, optical imaging, and multiphoton microscopy (MPM) [1,2]. In the latter group, two-photon excited fluorescence (TPEF) – a process in which two photons are absorbed simultaneously by a single fluorophore – is an important example of such a technique [3]. Currently state-of-art TPEF imaging systems primarily utilize commercially-available femtosecond Ti:Sapphire lasers delivering femtosecond pulses in the 700-800 nm window [4-7]. However, due to their architecture, Ti:Sapphire lasers have certain disadvantages hindering, or in many cases prohibiting real-life applications outside the laser physics laboratory. A Ti:Sapphire laser is based on a free-space resonator pumped by an expensive green laser (532 nm), besides active water cooling and an optical table are often required. As a result, the entire setup becomes complex, bulky and expensive. Working with Ti:Sapphire lasers might be bothersome, since they often require re-alignment or some adjustments to initiate a pulsed operation. Considering real-life applications in biological or medical research laboratories, clinics and hospitals, the TPEF imaging system should be as simple as possible, while daily operation of a Ti:Sapphire laser requires knowledge about laser optics. Moreover, intrinsically Ti:Sapphire lasers operate at relatively high pulse repetition frequencies ($f_{rep}$), often above 70 MHz. However, some applications require much lower $f_{rep}$, which implies using an external pulse-picker (based on bulk crystal) to reduce the repetition rate. A pulse-picker introduces unwanted chromatic dispersion and further increases the size, cost and complexity of the setup.

   Development of femtosecond lasers is often driven by needs of particular applications in biology and medicine. One such example is a need for low and adjustable $f_{rep}$ femtosecond sources, especially important for multiphoton microscopy. It has been shown that reducing the pulse $f_{rep}$ is beneficial for increasing TPEF signal during imaging of biological tissues. One example is imaging of skin and few other pigment-containing tissues, where thermal-mechanical damage can be mitigated by using a pulse picker [8]. The origin of this damage is strong one-photon absorption of melanin at the wavelength of excitation [9]. While picking pulses, the average laser power is reduced but the peak power is maintained. Moreover, it also allows sufficient time for heat dissipation between pulses. As a consequence, when allowed by biochemical damage threshold, it gives room to further increase the pulse peak power resulting in increased TPEF signal [10]. In another example it was shown that reduced $f_{rep}$ increases fluorescence yield through dark state relaxation. Increasing time between pulses to >1 μs ensures that transient molecular dark states with long lifetime will relax between two absorption events [11]. Finally, pulse pickers are commonly used in two-photon fluorescence lifetime imaging microscopy (FLIM) to adjust the pulse separation to the specific fluorophore's lifetime [12-14]. For those reasons ability to reduce and control the repetition rate is very important.

As an alternative to Ti:Sapphire lasers, compact frequency-doubled Erbium-doped fiber lasers can be used [15]. Fiber lasers, compared to solid state Ti:Sapphire lasers, are more compact and simple, robust, maintenance-free and cost effective. However, achieving few cycle pulses directly from Erbium-fiber oscillators is challenging. Sub-50 fs pulses were obtained from the setup utilizing free space optics [16], while in the case of all-fiber Erbium-lasers pulse duration was limited to ~50 fs by high order dispersion and nonlinear effects in optical fibers [17,18]. Another approach to achieve fs pulses is external amplification and compression, using non-standard fibers with precisely adjusted length, which allows to achieve even sub-20 fs pulses [19-23]. Pulse duration of 14 fs was achieved from an amplified carbon nanotube mode locked fiber laser; however, the setup includes at its output a highly nonlinear fiber (HNLF) and a prism pair [24]. Generation of 22.7 fs pulses with 2.8 nJ energy was reported in [25], by single stage soliton compression using a standard single-mode fiber (SMF) for both pre-stretching pulses from the oscillator, and their compression after amplification. All-fiber and polarization maintaining (PM) setup presented in [26] allowed to achieve 24 fs pulses with energy of 3 nJ. Presented setup utilizes only three commercially available PM fiber optic components and two types of PM fibers (also commercially available – SMF and active fiber), providing linear polarization of the output pulses and stable operation.

Femtosecond pulses at 1560 nm wavelength can be converted to ultrashort 780 nm pulses, to ideally match the two-photon absorption spectra of many fluorophores of interest, by adding a frequency-doubling stage for second harmonic generation (SHG). As an example, a high-power frequency-doubled laser with 520 fs pulses at 85 MHz repetition rate, resulting in pulse energy of 10.3 nJ and peak power of more than 19.8 kW was used for in-vivo TPEF spectroscopy and imagining [27]. However, the relatively long pulse duration and high repetition rate excludes such source from e.g. retinal imaging [4]. Another system for two-photon microscopy, generating 191 fs pulses with 156 MHz repetition rate and 1 W of average power was demonstrated in [28]. Generation of much shorter pulses - about 40 fs at 796 nm was presented in [29], which was achieved by converting 50 fs pulses at the central wavelength of 1600 nm by the SHG module. The average power of the signal at ~800 nm was 140 mW, corresponding to 3.5 nJ pulse energy at $f_{rep}$ of 40 MHz. The system does not contain any $f_{rep}$ control and requires three amplification stages. Moreover, configuration based on nonlinear frequency conversion creates a possibility for dual-wavelength operation for simultaneous two- and three-photon imaging in multimodal instrument. In [30], 80 fs pulses at 790 nm wavelength (61 mW of average power) were generated and used, simultaneously with residual, unconverted 80 fs pulses at 1580 nm (117 mW of average power). The residual 1580 nm pulses were used for third harmonic generation (THG) skin imaging, while 790 nm pulses for TPEF and SHG imaging. However, none of the configuration presented so far has provided an adjustable $f_{rep}$, required in many biomedical applications.

In this work we report a frequency-doubled Er-fiber laser system for TPEF imaging which addresses the main shortcoming of the previously reported setups: it enables pulse repetition rate tuning by means of a fiber-coupled pulse-picker integrated in the amplifier chain. The system delivers ultrashort, sub-65 fs pulses (60 fs at shortest) with >1.3 nJ energy preserved in the entire $f_{rep}$ tuning range. The applicability of the source for two-photon excited fluorescence microscopy was confirmed by imaging of *ex vivo* samples at various pulse repetition rates. We show that reducing the pulse repetition rate allows to increase fluorescence intensity from rat skin samples.

## 2. Design of the frequency-doubled Er-fiber laser source

Figure 1 shows the experimental setup of the frequency-doubled Er-doped fiber laser source. As a seed source, a ring cavity Er-doped fiber laser (EDFL) mode-locked via semiconductor saturable absorber mirror (SESAM) was used (Batop GmbH). The seed generated 325-fs pulses at 35.7 MHz repetition rate and average power of 1.1 mW at 1561 nm central wavelength. Next, the pulses were pre-amplified in an Erbium-doped fiber amplifier (EDFA) based on 1.5 m-long segment of highly-Erbium doped fiber (Liekki Er80-4/125-PM, EDF). After amplification, the pulses entered a pulse-picker (PP) based on a fast acousto-optic modulator (Gooch & Housego Fibre-Q 200 MHz, AOM). The details on the PP design were already presented in our previous work [31]. After $f_{rep}$ reduction, the pulses were amplified in a final amplification stage, based on the same EDF as in the previous stage. An additional band-pass filter (BPF) with 10 nm width (centered at 1560 nm) was placed before the power amplifier. The BPF limited the spectral width prior amplification and enabled achieving a clean pulse shape at the output, with minimized sidebands. The setup was entirely fiberized and spliced using a standard arc-fusion splicer. All fibers and components used in this setup were single-mode and polarization maintaining (PM-SMF), providing polarization stability which is crucial for frequency doubling. The entire system was built using only two types of fiber (the EDF and PM-SMF).

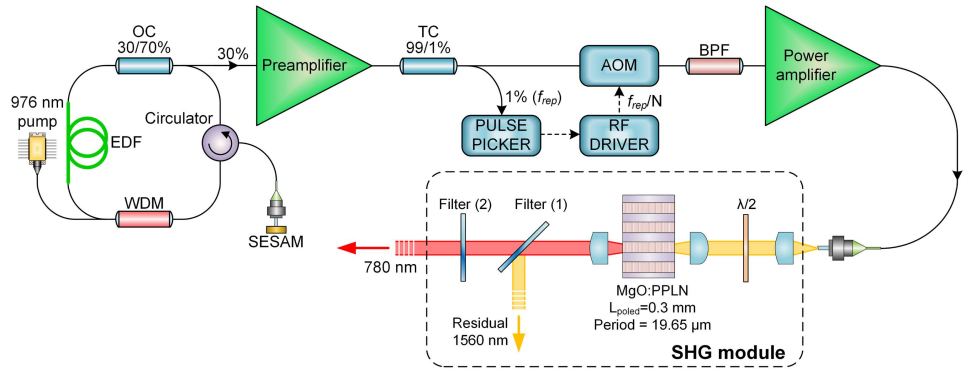

Fig. 1. Scheme of the frequency-doubled, fiber femtosecond laser: EDF – Erbium-doped fiber; OC - output coupler; WDM - wavelength division multiplexer; SESAM - semiconductor saturable absorber mirror; TC - tap coupler; AOM - acousto-optic modulator; BPF - bandpass filter.

The source is capable of generating pulses with duration down to 22 fs at any chosen repetition rate from 1 to 12 MHz. Figure 2 shows the optical spectrum recorded with an optical spectrum analyzer (Yokogawa AQ6370B, OSA) and the temporal intensity of the pulse together with the temporal phase, measured via frequency resolved optical gating technique (Mesa Photonics FS-Ultra2, FROG). Both measurements were taken at $f_{rep}$ set to 1.02 MHz. At this setting, the obtained average power was 6.5 mW which corresponds to pulse energy of approx. 6.37 nJ, more than twice than reported previously [25,26]. The energy is lower than in [29], however, our setup has only two amplification stages in comparison to three [29].

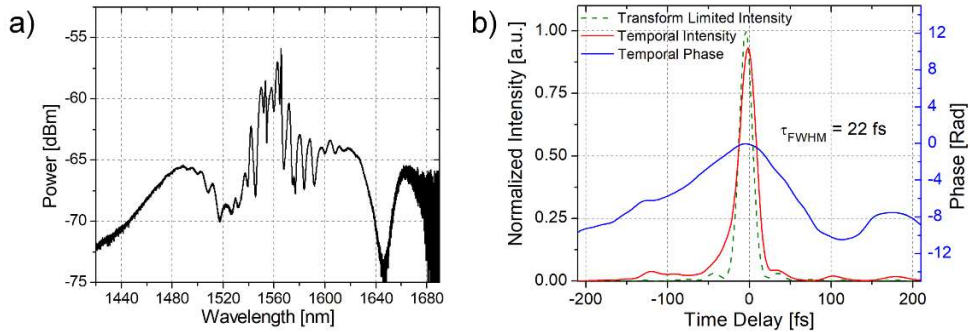

Fig. 2. (a) Optical spectrum after amplification at 1 MHz repetition rate and (b) FROG temporal intensity of the pulse (solid red line) together with the calculated transform-limited intensity (dotted green line) and temporal phase (solid blue line).

The ultrashort 1560 nm pulses were afterwards frequency-doubled in the SHG module. The amplifier output was collimated and focused by an achromatic lens (19 mm focal length) on a MgO:PPLN crystal (FSHNIR-ER, HC Photonics Corp.) with a 19.65 μm quasi phase matching period and 0.3 mm poling length ($L_{poled}$). The frequency-doubled beam was collimated and directed through two filters: a dichroic mirror which reflects the unconverted 1560 nm pump (Thorlabs DMSP1000, Filter 1) and a bandpass filter with cut-off at 700 nm (Thorlabs FELH0700, Filter 2) which blocks higher harmonics below 700 nm. The optical spectrum of the second harmonic and the pulse temporal intensity at $f_{rep}$ = 1.02 MHz are presented in Fig. 3(a) and (b), together with the measured and retrieved FROG spectrograms (c). The spectrum is centered at 782.6 nm with a full width at half maximum (FWHM) of 19 nm. The pulse duration is 60 fs and the average power (measured after both filters) equals to 1.4 mW, corresponding to a pulse energy of 1.37 nJ. It is noteworthy that the FROG measurement revealed a clean pulse shape without any pronounced sidebands, and a flat temporal phase.

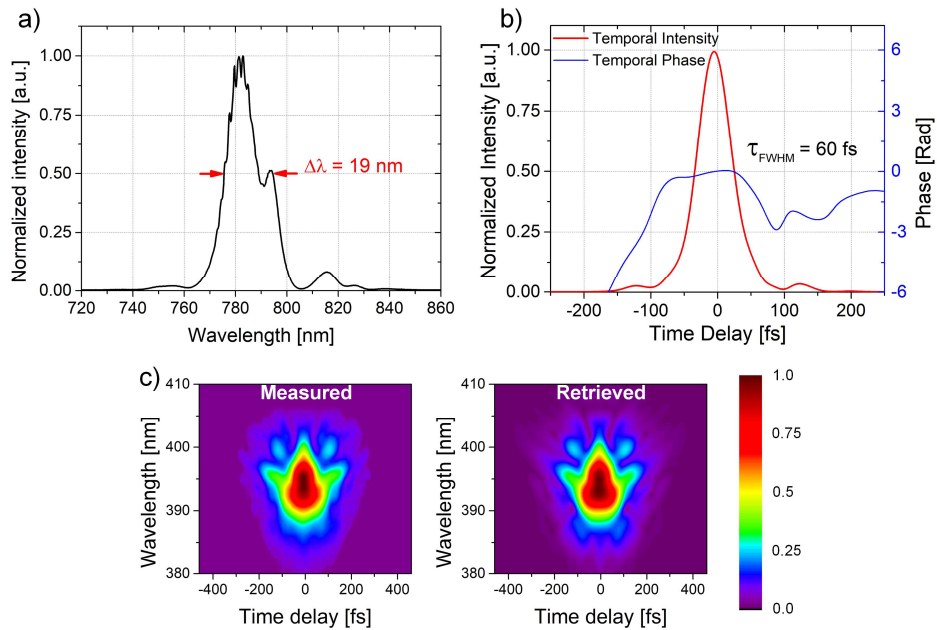

Fig. 3. (a) Generated second harmonic spectrum, (b) FROG temporal intensity of the 780 nm pulse (solid red line) together with the temporal phase (solid blue line), (c) measured and retrieved FROG spectrograms.

The used PP allows for easy repetition frequency tuning by dividing the fundamental $f_{rep}$ by a chosen factor [31]. Figure 4(a) presents the average output power and pulse duration characteristic of the second harmonic versus repetition rate. By increasing the $f_{rep}$ from 1.02 MHz up to 11.9 MHz we observed a linear increase of the average power (i.e. preserved pulse energy) and small changes of the pulse duration in the range of only 2 fs, which indicates that the source can be used in a wide range of frequencies with preserved pulse energy and duration, enabling optimization for specific experimental requirements (like lifetime of fluorophores, etc.). The output power and pulse energy remain stable over long periods of time. Figure 4(b) presents the pulse energy stability measured over 6 hours without any active stabilization (only passive thermal stabilization of the oscillator, amplifiers and PPLN crystal). It is worth noting that there is no warm-up time required to achieve the desired parameters - the system is fully turn-key operated, and the obtained stability is 0.7% rms.

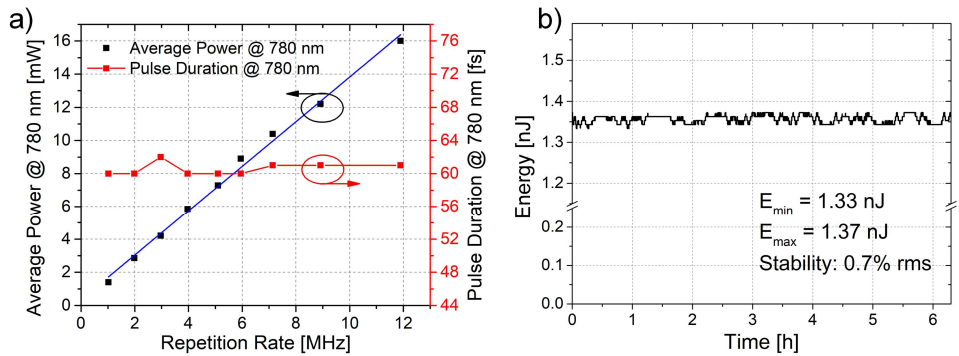

Fig. 4. (a) Measured average power and pulse duration of the 780 nm pulses as a function of the pulse repetition rate. (b) Long-term energy stability measurement over a period of 6 hours, indicating stability of 0.7% rms.

## 3. Application to TPEF microscopy

To verify the laser applicability to nonlinear imaging and confirm the expected effect of decreasing the pulse repetition rate, the laser was coupled to a home-build TPEF microscope schematically shown in Fig. 5(a). A variable neutral density (ND) filter placed in front of the microscope allows to adjust the average power level at the sample. Next, the beam is directed to galvanometer scanning system (Thorlabs GVS002), then through a scan lens (Thorlabs AC254-50-B-ML) and tube lens (Thorlabs AC254-100-B-ML) to a microscope objective (Olympus LUCPlanFL 40x/0.6 NA). The microscope objective was slightly underfilled to compromise the resolution and power efficiency. The emitted fluorescence is collected in epi-mode in a de-scanned manner. We use a dichroic mirror (Semrock HC 705 LP) to separate fluorescence from excitation light and a set of two additional optical filters (Semrock 694/SP BrigthLine HC). The fluorescence light is collected and processed by a photon counting GaAsP photomultiplier module (Hamamatsu H7422-40P), photon counting module (Hamamatsu C9744), data acquisition card (National Instruments USB-6212), and a custom-written LabView software. The images were acquired with 40 μs dwell time and 512x512 pixels, resulting in 0.1 frames/s. In order to improve signal to noise ratio 10 frames were typically collected for averaging. The fluorescence intensity is quantified as a mean count of photons per pixel (mpc) within the image.

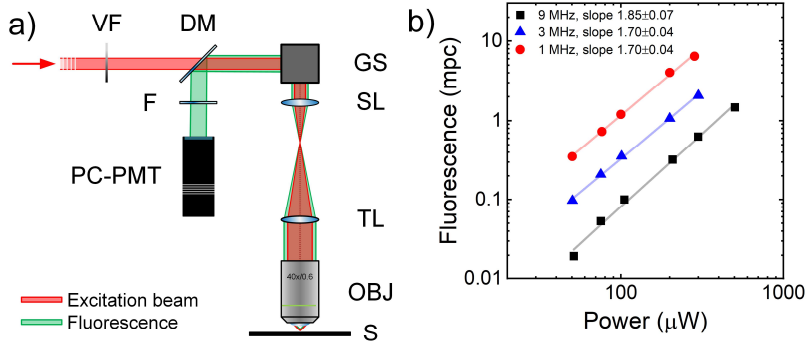

Fig. 5. (a) Experimental setup of a home-built TPEF microscope. VF – variable neutral density filter, DM – dichroic mirror, GS – galvanometer scanning unit, SL – scan lens, TL – tube lens, OBJ – objective lens, S – sample, F – cut-off filter, PC-PMT – photon counting photomultiplier. (b) Fluorescence of test target in mean photon counts per pixel (mpc) as a function of excitation average power for 1, 3, and 9 MHz pulse repetition rates. Slopes of linear regression lines through the data points on the log-log plot indicate two-photon process for all pulse repetition rates.

The fluorescence intensity depends on the number of absorbed photon pairs per second ($N_a$) in two-photon process [8,10]:

$$N_a \propto \frac{1}{f_{rep}} \cdot \frac{\delta}{A^2 \tau} \cdot P_{avr}^2, \tag{1}$$

where $f_{rep}$ is the pulse repetition rate, $\delta$ is the two-photon absorption cross-section, $A$ is the spot size of the beam in the focus, $\tau$ is the pulse duration (in the sample plane), and $P_{avr}$ is the average excitation power. According to this relationship, it is possible to increase the fluorescence signal by decreasing the pulse repetition rate, if the average power at the sample is maintained at the same, unchanged level. We tested this relation by measuring the fluorescence of a fluorescence test target (Edmund Optics, #57-894) as a function of power for pulse repetition rates of 9, 3, and 1 MHz. In all cases slopes of linear regression in log-log plot are ~1.8 (see Fig. 5(b)), indicating a second-order nonlinear process with small contribution of linear absorption possibly originating from a broad absorption spectrum of the test target. It is also visible that the fluorescence intensity gradually increases with decreased pulse repetition rate (with respect to 9 MHz – 2.73 times for 3 MHz and 9.34 times for 1 MHz repetition rate).

Next, we applied the developed laser and two photon microscope for *ex vivo* imaging the cross-section of unstained albino rat skin sample. Figure 6 (a)-(f) shows six representative TPEF images recorded at various pulse repetition rates and two average power levels. The variable ND filter was adjusted to obtain desired average excitation power for a given pulse repetition rate. Each image is showing the same location within the sample displaying oval structures of hair follicles. It is apparent that reducing the repetition rate to 1 MHz results in brighter image. For quantitative analysis, we recorded the fluorescence intensity as a function of power for 9, 3, and 1 MHz repetition rates and plotted in Fig. 6(g). A quadratic function ($y = C \cdot x^2$) was fitted to each measurement series. We obtained 3.56 times higher value of $C$ parameter for 3 MHz repetition rate with respect to 9 MHz, while 9.40 times higher when reducing to 1 MHz. This means that it is possible to obtain the same fluorescence intensity with 1 MHz excitation using $9.40^{1/2} = 3.07$ times less power when compared with 9 MHz excitation, which reduces the thermal damage probability [8]. Minor deviation from theoretically expected values of $C$ parameters (3 and 9 for 3 MHz and 1 MHz pulse repetition rates, respectively) is caused by small differences in pulse duration for different repetition rates, as seen in Fig. 4. For further analysis we recorded the fluorescence intensity as a function of repetition rate, without adjusting the variable ND filter. In this experiment a constant pulse energy of 65 pJ in the sample plane is maintained, while average power increases with repetition rate. The fluorescence intensity grows linearly, as expected and is shown in Fig. 6(h). Figure 6(i) shows TPEF image of the sample obtained with 1 MHz and 440 µW of excitation power in larger field of view of 600x600 µm. Summarizing, this demonstrates that the laser is performing according to expectations and can be used for imaging with various pulse repetition rates.

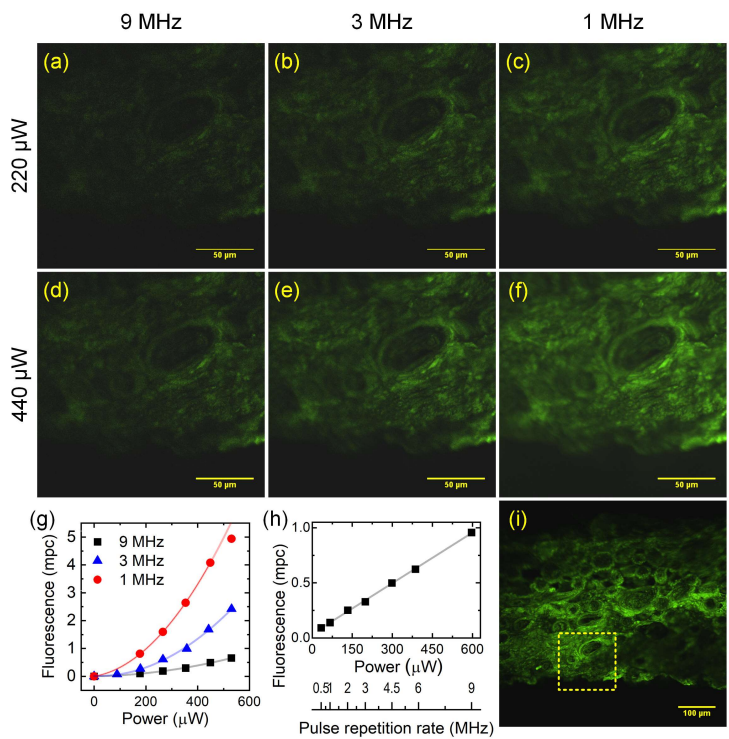

Fig. 6. Reducing the pulse repetition rate while maintaining the average excitation power allows to increase fluorescence intensity from *ex vivo* rat skin samples. (a)-(c) Imaging with 220 µW excitation power and 9, 3, and 1 MHz repetition rates. (d)-(f) Imaging with 440 µW excitation power and 9, 3, and 1 MHz repetition rates. (g) Fluorescence intensity of *ex vivo* rat skin in mean photon counts per pixel (mpc) as a function of excitation average power for 1, 3, and 9 MHz repetition rates; the last point in 1 MHz data series deviates from fit function due to saturation of photon counting electronics. (h) Fluorescence intensity of rat skin as a function of pulse repetition rate (excitation power) and maintained pulse peak power. (i) Image with larger field of view obtained with 1 MHz repetition and 440 µW excitation power; selected ROI indicates sample region where (a)-(f) images were taken.

As another example, various biological samples were imaged with 1 MHz pulse repetition rate excitation using their endogenous autofluorescence. Figure 7(a) shows the cross-section of a frog liver on prepared microscope slide obtained with 380 µW excitation power. A characteristic structure of hepatocyte cells is visible. Figures 7(b) and (c) show the autofluorescence image of freshly picked plant leaf of *Chamaedorea elegans* and *Epipremnum scindapsus*, respectively. In both cases the excitation power was less than 200 µW in the sample plane. The fluorescence is originating from chlorophyll stored in the chloroplasts. All in all, it demonstrates that it is possible benefit from adjustable repetition rate of the laser source for low excitation power imaging of biological samples.

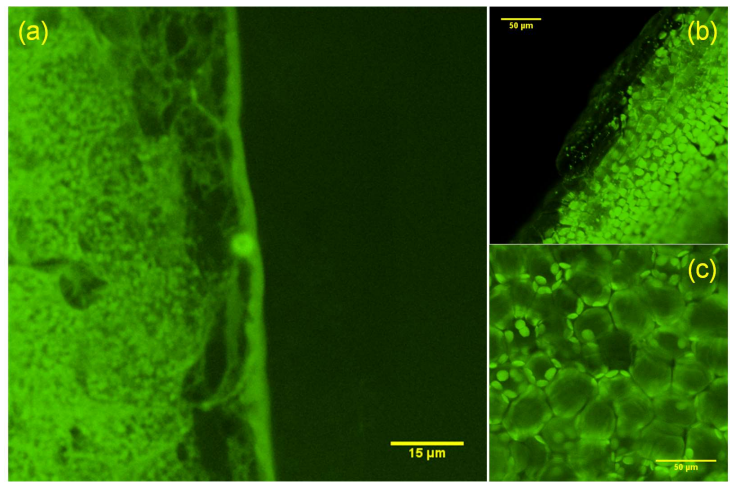

Fig. 7. Exemplary TPEF images of unstained biological samples obtained with 1 MHz pulse repetition rate. (a) *Ex vivo* frog liver cross-section imaged with 380 μW excitation power. (b) *Chamaedorea elegans* leaf imaged with 115 μW excitation power. (c) *Epipremnum scindapsus* leaf imaged with 154 μW excitation power.

## 4. Summary

In conclusion, we report an exceptionally simple and compact system for low power, efficient two photon excited fluorescence microscopy with a fiber laser source of ultrashort pulses at 780 nm wavelength. The light source is based on a frequency-doubled Er-doped fiber laser with built-in pulse-picker for easy repetition rate adjustment. New set-up enables easy repetition rate tuning in the range of 1 to 12 MHz, with preserved sub-65 fs pulse duration (60 fs at shortest) and energy exceeding 1.3 nJ. The performed experiments at various biological samples confirm that the light source meets the requirements of two-photon excited fluorescence microscopy, and it is possible benefit from adjustable pulse repetition rate. As an example, we showed that it is possible to increase the TPEF signal by 9.4 times by reducing the repetition rate from 9 to 1 MHz with maintained average power level. Moreover, our laser system can be an excellent platform for development of adaptive excitation source for high-speed imaging of neuronal activity [32]. We believe that such robust and reliable fiber-based sources can replace Ti:Sapphire lasers as excitation sources in outside-laboratory applications.

**Funding**

Foundation for Polish Science (First TEAM/2017-4/39, TEAM TECH/2016-3/20, The International Centre for Translational Eye Research (MAB/2019/12)). European Union's Horizon 2020 (No. 666295), and from the financial resources for science in the years 2016-2019 awarded by the Polish Ministry of Science and Higher Education for the implementation of an international co-financed project.

**Acknowledgments**

We thank Maciej Popenda and Bogusław Szczupak from Wrocław University of Science and Technology for providing the 780 nm pulse autocorrelator for preliminary measurements. We acknowledge the invaluable help of Piotr Ciąćka in LabView programming, also Jadwiga Milkiewicz and Łukasz Kornaszewski (Institute of Physical Chemistry PAS) for helpful discussions. We thank Hubert Doleżyczek from Nencki Institute of Experimental Biology for preparation of rat skin samples.

## Disclosures

The authors declare no conflicts of interest.